# Network congestion control using NetFlow

*Maxim A. Kolosovskiy*   *Elena N. Kryuchkova*

*Altai State Technical University, Russia*

*Abstract*

*The goal of congestion control is to avoid congestion in network elements. A network element is congested if it is being offered more traffic than it can process. To detect such situations and to neutralize them we should monitor traffic in the network. In this paper, we propose using Cisco's NetFlow technology, which allows collecting statistics about traffic in the network by generating special NetFlow packets. Cisco's routers can send NetFlow packets to a special node, so we can collect these packets, analyze its content and detect network congestion. We use Cisco's feature as example, some other vendors (Juniper, 3COM, Alcatel, etc.) provide similar features for their routers. We also consider a simple system, which collects statistical information about network elements, determines overloaded elements and identifies flows, which congest them.*

# 1. Introduction

The goal of **congestion control** is to avoid congestion in network elements. A network element is congested, if it is being offered more traffic than it can process. Congestion control is control of resources: routers CPUs, bandwidth of links, routers memory, etc.

If we don't control our network, there is potential for a serious trouble. When some network element becomes congested, it processes traffic very slowly and some packets are lost. Therefore, users don't receive expected packets (or conformation of delivery) in the time limit. Users begin to resubmit packets and new packets cause further congestion. Such situation is called **congestive collapse**. Congestive collapse was explored as possible problem in 1984 by John Nagle [1]. First well-known congestive collapse happened with NSFNet in 1986: capacity of backbone dropped from 32 kbit/s to 40 bit/s. First congestion control algorithm was introduced by Van Jacobson in 1987 [2].

If we can't offer high-speed service to all users, we should restrict consumption of resources by users to avoid congestion. So, all users get satisfactory service continuously. If we permit unlimited using of resources, it can cause congestive collapse and all network services become unavailable.

**Congestion control system** usually works in the following way. The system monitors various factors (e.g. router's CPU occupancy, link occupancy, percent of delivered packets, messaging delays, etc.). Based on this information the system detects possible congestion. If the system has detected an incipient congestion, it restricts traffic rates in some network elements and continues to monitor the state of the network. This activity reduces traffic through these elements. When elements will have been unloaded, the system will restore normal traffic rates in these elements.

Thus, congestion control system must have opportunity to monitor the state of the network. A traditional approach is using SNMP. SNMP-approach is very limited: we can't get detailed information about the network. For detailed information, we can use special features (e.g. NetFlow, IPFIX, jFlow, sFlow, NetStream). The main idea is generating packets, which contain information about router's active flows. The features permit to decompose network flows according to various attributes (IPs, protocols, interfaces, ports, etc.).

You can find information about congestion control in [7, 8]. Many RFCs have been written about congestion control, section References contains some of them [3-6].

## 2. Fundamentals of NetFlow technology

**NetFlow** is Cisco's technology, which allows collecting information about network traffic [9]. NetFlow technology includes three parts:
- **NetFlow exporter** enables to generate UDP packets, which contain information about flows passing through the router, and send information to special node (NetFlow collector). Cisco's routers can be exporters.
- **NetFlow collector** saves received packets to the data storage;
- **Analyzer** is special software, which analyses received data.

A router considers traffic as a collection of flows and stores a cache entry for each active flow. A router inspects each received packet and creates a new cache entry (if the packet doesn't belong to active flows) or updates one of existing entries (if the packet belongs to an existing flow). Two packets are from one flow, if they have the same following fields:
- Source IP address;
- Destination IP address;
- Source port number;
- Destination port number;
- Layer 3 protocol type;
- Type of service byte;
- Router or switch interface.

These fields are called **key-fields**; other fields are called **non-key-fields**.

When a flow becomes inactive, a router places its cache entry to NetFlow packet. A router considers a flow inactive, if a router doesn't receive any packets of the flow (by default: 15 sec). A router also considers a flow inactive, if a flow is long lived (by default: 30 min). UDP doesn't guarantee reliable service, so datagrams may arrive out of order, appear duplicates and some datagrams may be lost. To correct such errors a packet contains a special field: number of flows, which were inspected by the router. Using this field we can reject duplicates and process datagrams in the correct order.

Each NetFlow packet consists of a header and a sequence of flow records. A record contains information about one flow (IP addresses, ports, number of packets, etc). A header contains version of NetFlow, number of flows in this packets (1-30), time since the router booted, etc.

**Flexible NetFlow** overcomes limitations with key-fields: we can specify own set of key-fields [10]. So we receive only interesting information, we can aggregate all flows in groups before exporting to the collector (for example, if we would like to analyze protocol usage, then we specify single key-field – protocol type), Flexible NetFlow introduce a collection of new available fields (for example, IPv6, information from layer 2 to layer 7).

Versions 5 and 9 are most popular versions of NetFlow [11]. **IPFIX** (IP Flow Information Export) is version 9 standardized by IETF [12]. We use Cisco's feature as example, some other vendors (Juniper, 3COM, Alcatel, etc.) provide similar features for their routers. You can find more information about NetFlow on Cisco's official site www.cisco.com.

NetFlow provides only collecting of statistical information about network. We can apply this information to detect congestion, to mining dependency between flows, to detect anomalies and viruses, to optimize the utilization of network resources, to monitor application and network usage, etc. (as examples [13-15]). These problems are interrelated. For example, some viruses initiate network congestion, so congestion control system can work as indicator of active viruses; knowledge about network dependencies helps to understand causes of congestion.

## 3. Using NetFlow for congestion control

A traditional method of performance monitoring is using SNMP (Simple Network Management Protocol). SNMP permits to monitor bandwidth, so we can use information obtained by SNMP for capacity planning. However, it is not enough to understand how well the network supports the business. More detailed knowledge of flow is extremely important in networks today. Understanding, which applications and IP addresses are generating the traffic, is priceless.

Using NetFlow technology, we obtain granular information about usage of network elements:

- we can monitor number of bytes and number of packets, which are processed by each router and by each link (using corresponding fields in NetFlow packets);
- we can analyze, which flows stress certain routers and links (using key-fields of active flow: source and destination IP, ports, protocol, ToS, interface);
- we can use timestamps fields to build historical usage trends (start of the flow, last received packet of the flow). Historical trends can be built for different periods of time: hours, days, weeks, months, etc.;
- we know paths used for packet delivering (field "Next Hop");
- we can collect statistics only about interested interfaces;
- we can observe interaction between autonomous systems, group of addresses, subnets.

In one word, all fields in NetFlow packets can be useful. Extended collection of Flexible NetFlow fields make this technology more attractive.

So, we can obtain information about existing (or incipient) congestion. If we've obtained necessary information about network, we can analyze the state of the network to find overloaded network elements, which work as "bottleneck". Next step is to restrict traffic in these elements (by reducing bandwidth of some routers or links) or to re-distribute the part of traffic between other network elements (by changing routing tables in routers). On the contrary, if the system detects that some network elements work with too low rates, the system increases the load of these elements.

Congestion control system can work in different ways:

- **Off-line monitoring system.** The system can analyze collected historical information about network and report about overloaded segments of network. To unload these segments the system provides information about causes of existing problems. If we have collected information for a long period, we plan changes of our network (e.g. capacity and topology planning).

- **On-line monitoring system.** The system analyzes and collects information simultaneously (instead of previous case, where we analyze collected information). We obtain information about the network online, so we can manually change the network state.
- **Controlling system.** The system can change parameters and state of the network online. The system usually changes routing tables and traffic rates. Instead of previous case this system makes changes in automatically mode.

Types of system are listed in increasing order of efficiently. First system can't prevent serious congestion problems; we can only analyze causes of these problems. Using the next system, we can detect incipient congestion, but we should make changes manually and if the network is large, it may be inconvenient. Last system is the most effective one, but more complex; this system works automatically.

A system can be organized as a collection of alerts. Each alert corresponds to a resource of a network element. If usage of the resource exceeds assigned threshold, the system signalizes about incipient congestion.

A necessary part of the congestion control system is data storage. Ordinary files can be used, but for large networks it should be a database.

# 4. Example of congestion control system

We would like to discuss simple congestion control system in Java. Using of our system consists of the following steps:

1. Configuring hosts in the network to send NetFlow v5 packets to our NetFlow collector.
2. Collecting information about network flows for a long period. The collector saves necessary fields of packets to a file (source and destination IPs and ports, timestamps, protocols, number of Layer 3 bytes, etc.).
3. Building some tables for analyzing the state of the network.
4. Analyzing tables and changing network parameters.

To analyze data we build the following statistical tables (to make examples more clear we use NetFlow packets, which are generated artificial):

### A. The load of hosts

```
IP                  Total

10.1.12.2           2763.8 MB [6%-22%-23%-40%-3%-4%]

10.1.12.10          2777.6 MB [1%-22%-21%-33%-16%-3%]

10.1.12.12          4523.3 MB [5%-13%-22%-20%-13%-25%]
```

Using the table we can estimate the load of each host. We can replace our routers corresponding to their load. For each host we show its IP address, total traffic and a distribution of the traffic over different periods of a day (we assume that a day consists of six 4-hour periods). Hosts are sorted in increasing order by total traffic.

### B. The load of links

```
            10.1.12.8 => 10.1.12.6                Total: 288.7 MB

    10.1.12.1:29750   ->  10.1.12.6:12352          52% (151.1 MB)
    10.1.12.13:29792  ->  10.1.12.4:23725          42% (121.9 MB)
    10.1.12.8:20644   ->  10.1.12.14:16906          5% (15.7 MB)
```

The example shows statistic for a link. We can see conversations, which are using maximal bandwidth. In this example we see, that first two conversations load the link most of all, so if we want to unload the link, we should change the path of one of these conversations. If we change the path of last conversation, it doesn't change the load of the link essentially.

### C. The most unloaded links

```
1.      10.1.12.15 => 10.1.12.11            4.2 MB

2.       10.1.12.5 => 10.1.12.11           11.1 MB

3.       10.1.12.5 => 10.1.12.4            11.6 MB

4.       10.1.12.4 => 10.1.12.16           11.6 MB

5.       10.1.12.6 => 10.1.12.14           15.7 MB
```

We can use these links to unload other links.

**D. Conversations**

```
10.1.12.14:28542  ->  10.1.12.3:29828
      10.1.12.6  => 10.1.12.8         48.7 MB   (of 811.0 MB)
      10.1.12.7  => 10.1.12.12        48.7 MB   (of 416.5 MB)
      10.1.12.8  => 10.1.12.7         48.7 MB   (of 457.6 MB)
      10.1.12.14 => 10.1.12.6         48.7 MB   (of 481.1 MB)
      10.1.12.12 => 10.1.12.3         48.7 MB   (of 1104.3 MB)
```

*Conversation* is defined by four parameters: source and destination IPs and ports. For each conversation, we build the list of links, which are used for that conversation (in parentheses we show total traffic in the link).

**E. Interactions with other hosts and protocol distribution**

```
10.1.12.3
 10.1.12.14       62.3 MB (68%)
 10.1.12.4        13.5 MB (14%)
 10.1.12.16       10.4 MB (11%)
 10.1.12.6         4.8 MB (5%)
Total traffic:    91.0 MB
Protocols: TCP - 77%, UDP - 11%, Other - 11%
```

For each host we display the list of hosts, which interact with that host. We also display total traffic for each interaction. Last line contains protocol distribution for that host.

**F. Input and output traffic**

```
*** 10.1.12.1 ***
IN       165.8 MB [0%-0%-62%-2%-0%-34%]
OUT      292.2 MB [0%-4%-47%-47%-0%-0%]
IN&OUT   458.0 MB [0%-2%-53%-31%-0%-12%]
```

For each host we display input, output and total traffic. We also display distribution of the traffic over time (day is divided into six 4-hour periods).

**G. Active ports of host**

```
10.1.12.2
 :4157    473.9 MB (33%)
 :31890   438.1 MB (31%)
 :15681   237.1 MB (16%)
 :9015    149.3 MB (10%)
 :21435    64.6 MB (4%)
 :22222    27.4 MB (1%)
 :9390     10.2 MB (0%)
Total:   1400.6 MB
```

Active ports of the host (with total traffic). Each network service is assigned to one or more ports; so we can estimate how many traffic the service produces and consumes.

**Use case**: We are not satisfied by working of the service, which are located on *10.1.12.7* and uses port *32001*. We find hosts, which interact with *10.1.12.7* and find conversations, which are using port *32001*. So, we know links and routers, which support this conversation. We analyze the load of these elements and find "bottlenecks", which delay the service. We know conversations, which load bottlenecks. Finally, we change parameters of some network elements to unload bottlenecks.

Thus, these tables permit to analyze the state of the network, to detect possible congestion and to make a decision about changes. We show elementary examples. These examples can be combined to produce tables that are more complex.

## 5. Conclusion

Network congestion is an actual problem today. We have considered using NetFlow for network congestion control, because NetFlow is very powerful tool for collecting information about network traffic.

To optimize network we can change routing tables on routers, we can replace some of network elements (routers and links) and we can change the topology of the network. Obtained data allow distributing application between hosts in optimal way (for example, using information about conversations we can minimize traffic in the network).

Suitable data storage is very important for such systems. We use ordinary files, but it is the simplest way. It is not satisfactorily for large data streams, so we should use another data storage in our system. There are some novel solutions to support high-speed incremental collecting and analyzing traffic information (e.g. *aggregation databases* [16], *sketches* [17]). These solutions use special properties of flow-based traffic information to make data processing more effectively.

Using our system network administrator obtains information about overloaded (and unloaded) elements, but the system doesn't propose explicit actions to unload congested elements. Thus, the next step is developing algorithms, which analyze collected information and find a solution (for example, the system proposes optimal way to redistribute traffic).